\documentclass[preprint,superscriptaddress,preprintnumbers]{revtex4}

\usepackage{graphicx}
\def\Re{{\cal R}e}
\def\ubar{\bar{u}}
  
\newcommand{\fslash}[1]{\mbox{$\!\not\!#1$}}

\begin{document}

\title{Two-photon exchange corrections to the pion form factor}

\author{P. G. Blunden}
\affiliation{Department of Physics and Astronomy,
    University of Manitoba, Winnipeg, MB, Canada R3T 2N2
    \footnote{Permanent address}}
\affiliation{Jefferson Lab, Newport News, Virginia 23606, USA}
\author{W. Melnitchouk}
\affiliation{Jefferson Lab, Newport News, Virginia 23606, USA}
\author{J. A. Tjon}
\affiliation{Physics Department, University of Utrecht, The Netherlands}

\begin{abstract}
We compute two-photon exchange corrections to the electromagnetic form
factor of the pion, taking into account the finite size of the pion.
Compared to the soft-photon approximation for the infrared divergent
contribution which neglects hadron structure effects, the corrections
are found to be $\lesssim 1\%$ for small $Q^2$ ($Q^2 < 0.1$~GeV$^2$),
but increase to several percent for $Q^2 \gtrsim 1$~GeV$^2$ at extreme
backward angles.
\end{abstract}

\maketitle

As the lightest bound state of quarks, the pion plays a unique role
in QCD.  On the one hand, its anomalously small mass leads to the
identification of the pion with the pseudo-Goldstone mode of dynamical
chiral symmetry breaking in QCD.  On the other, scattering experiments
reveal a rich substructure which can be best described in terms of
its quark constituents.  The most basic observable which characterizes
the structure of the pion is its electromagnetic form factor,
$F_\pi(q^2)$, where $q^2$ is the four-momentum transfer squared.

The extractions of the pion form factor in the space-like region
($Q^2\equiv -q^2>0$) from measurements of the pion electroproduction
reaction $e p \to e \pi^+ n$ have recently provided data on the $Q^2$
dependence of $F_\pi$ up to values of $Q^2 \sim 2.5$~GeV$^2$
\cite{FpiJLab}, and higher-$Q^2$ measurements are planned to
$Q^2 \sim 6$~GeV$^2$.  The main uncertainty in the extraction of
$F_\pi$ is the need to use models to extrapolate the longitudinal
electroprodution cross section to the physical pion mass assuming
pion pole dominance of the $t$-channel process \cite{FpiJLab}.
The pion electroprodution experiments complement low-$Q^2$ data
obtained by scattering pions from the electrons of a hydrogen target.

As is standard in most electromagnetic scattering analyses, the pion
form factor has been obtained from data assuming the validity of the
one-photon exchange, or Born, approximation.
Recently the accuracy of one-photon exchange approximation has been
called into question by the observation of a large discrepancy between
the proton electric to magnetic form factor ratio in measurements
using Rosenbluth separation and polarization transfer \cite{Jones}.
A number of detailed studies have demonstrated that these can be
mostly understood once radiative corrections arising from two-photon
exchange are included, in particular those associated with hadron
finite-size effects.  These findings have prompted exploration of
the significance of two-photon exchange in other reactions
(see Refs.~\cite{review} for reviews).

In this paper we investigate the role of two-photon exchange (TPE)
in electromagnetic scattering from the pion.  We use the methodology
developed for the application of TPE to scattering from the nucleon
\cite{BMT03,BMT05,KBMT05,KB07}, suitably modified to the scalar case.
The analysis of TPE from a spin-0 target is, in fact, considerably
simpler than that for spin-$\frac{1}{2}$ targets.

For the elastic electron--pion scattering process, we follow the
notation of Refs.~\cite{BMT03,BMT05} and define the momenta of the
initial electron and pion as $p_1$ and $p_2$, and of the final electron
and pion as $p_3$ and $p_4$, respectively,
$e(p_1) + \pi(p_2) \to e(p_3) + \pi(p_4)$.
The matrix element of the pion current is given by
\begin{eqnarray}
\langle \pi(p_4) | J^\mu(0) | \pi(p_2) \rangle
&=& (p_2 + p_4)^\mu\, F_\pi(q^2)\, ,
\end{eqnarray}
where $q^2 = (p_4-p_2)^2$.
In the one-photon approximation the scattering amplitude is given by
\begin{eqnarray}
{\cal M}_0
&=& {e^2 \over q^2} \bar u_e(p_3) \gamma_\mu u_e(p_1)\,
    (p_2+p_4)^\mu\, F_\pi(q^2)\, .
\end{eqnarray}
In the target rest frame of the pion the Born cross section can then
be written simply as
\begin{eqnarray}
{ d\sigma_0 \over d\Omega } 
&=& \sigma_{\rm Mott}\, F^2_\pi(q^2)\, ,
\end{eqnarray}
where
\begin{eqnarray}
\sigma_{\rm Mott}
&=& { \alpha^2 E_3 \cos^2(\theta/2) \over  
      4 E_1^3 \sin^4(\theta/2) }
\end{eqnarray}  
is the Mott cross section for electron scattering from a point
particle, with $E_1$ and $E_3$ the initial and final electron
energies, and $\alpha = e^2/4\pi$ the electromagnetic fine structure
constant.

Including ${\cal O}(\alpha)$ radiative corrections leads to a
modification of the Born cross section arising from vertex corrections,
vacuum polarization, inelastic bremsstrahlung, and two-photon exchange.
As discussed in Refs.~\cite{BMT03,BMT05}, only the latter lead to a
dependence on the scattering angle, or equivalently on the virtual
photon polarization parameter
$\varepsilon = \left(1 + 2 (1+\tau) \tan^2{(\theta/2)}\right)^{-1}$,
where $\tau = Q^2/4m_\pi^2$, and $m_\pi$ is the pion mass.
While the scattering angle naturally depends on the reference frame,
we can more generally express $\varepsilon$ in terms of Lorentz
invariants as
\begin{eqnarray}
\varepsilon &=&
{\nu^2 - \tau (1+\tau) \over \nu^2 + \tau (1+\tau)}\, ,
\end{eqnarray}
where $\nu \equiv p_1\cdot p_2/m_\pi^2 - \tau$.
In the target rest frame we have $p_1 \cdot p_2 = E_1 m_\pi$.

The total TPE amplitude, including the box and crossed-box
diagrams, has the form
\begin{eqnarray}
{\cal M}_{\gamma\gamma}
&=& i e^4 \int {d^4 k_1\over (2\pi)^4}
{L_{\mu\nu} H^{\mu\nu}\over (k_1^2-\lambda^2)(k_2^2-\lambda^2)}\, ,
\label{eq:Mgg}
\end{eqnarray}
where $k_1$ and $k_2$ are the momenta of the virtual photons,
with $k_1+k_2=q$.
The parameter $\lambda$ is introduced as an infinitesimal photon mass
in the photon propagators to regulate the infrared (IR) divergences.
The leptonic tensor $L_{\mu\nu}$ is given by
\begin{eqnarray}
L_{\mu\nu}
&=& \ubar_e(p_3)
\left[ \gamma_\mu
       {\fslash{p_1}-\fslash{k_1} + m_e \over (p_1-k_1)^2 - m_e^2}
       \gamma_\nu\
    +\ \gamma_\nu
       {\fslash{p_1}-\fslash{k_2} + m_e \over (p_1-k_2)^2 - m_e^2}
       \gamma_\mu
\right] u_e(p_1)\, ,
\end{eqnarray}
where $m_e$ is the electron mass.
The hadronic tensor $H^{\mu\nu}$ in principle contains contributions
from all excitations of the initial state.
In practice we approximate this by the pion elastic contribution
\begin{eqnarray}
H^{\mu\nu}
&=& F_\pi(k_1^2)\, F_\pi(k_2^2)\,
   {(2 p_2+k_1+q)^\mu (2 p_2+k_1)^\nu \over (p_2+k_1)^2-m_\pi^2}\, .
\end{eqnarray}
The TPE contribution to the cross section is then given by the
interference of the TPE amplitude ${\cal M}_{\gamma\gamma}$ and
the Born amplitude ${\cal M}_0$.
This can be parametrized in terms of a multiplicative correction
$1+\delta$, where
\begin{equation}
\delta = { 2 \Re \{ {\cal M}_0^*\, {\cal M}_{\gamma\gamma} \}
           \over |{\cal M}_0|^2}\, .
\label{eq:delta}
\end{equation}
The pion form factor is then modified according to
\begin{eqnarray}
F^2_\pi(q^2) &\to& F^2_\pi(q^2) (1 + \delta)\, .
\end{eqnarray}  

Experimental analyses of electromagnetic form factor data typically
use radiative corrections computed by Mo \& Tsai (MT) in the soft-photon
approximation \cite{MT69}, in which hadronic structure effects are
neglected. The TPE corrections are approximated by taking only the
IR-divergent contribution at the photon poles, setting $k_1\to 0$
and $k_2\to 0$ in the numerator of Eq.~(\ref{eq:Mgg}).
In this approximation ${\cal M}_{\gamma\gamma}$ becomes proportional
to the Born amplitude ${\cal M}_0$, and the corresponding correction
$\delta$ to the Born cross section is independent of hadronic structure
(or indeed of the type of hadronic target).
Mo \& Tsai approximate the remaining loop integration by further
reducing it to a 3-point function
$K(p_i,p_j) = p_i\cdot p_j\,\int_0^1 dy\,\ln{(p_y^2/\lambda^2)}/p_y^2$,
where $p_y=p_i y + p_j (1-y)$, with the total box plus crossed-box
contribution given by
\begin{equation}
\delta_{\rm IR}^{(\rm MT)}
= -2{\alpha\over \pi} \left[ K(p_1,p_2) - K(p_3,p_2) \right]\, .
\label{eq:deltaIRMT}
\end{equation}
The logarithmic IR singularity in $\lambda$ is exactly canceled by
a similar singularity arising from the bremsstrahlung correction
involving the interference between real photon emission from the
electron and from the pion.

\begin{figure}[t]
\includegraphics[height=6cm]{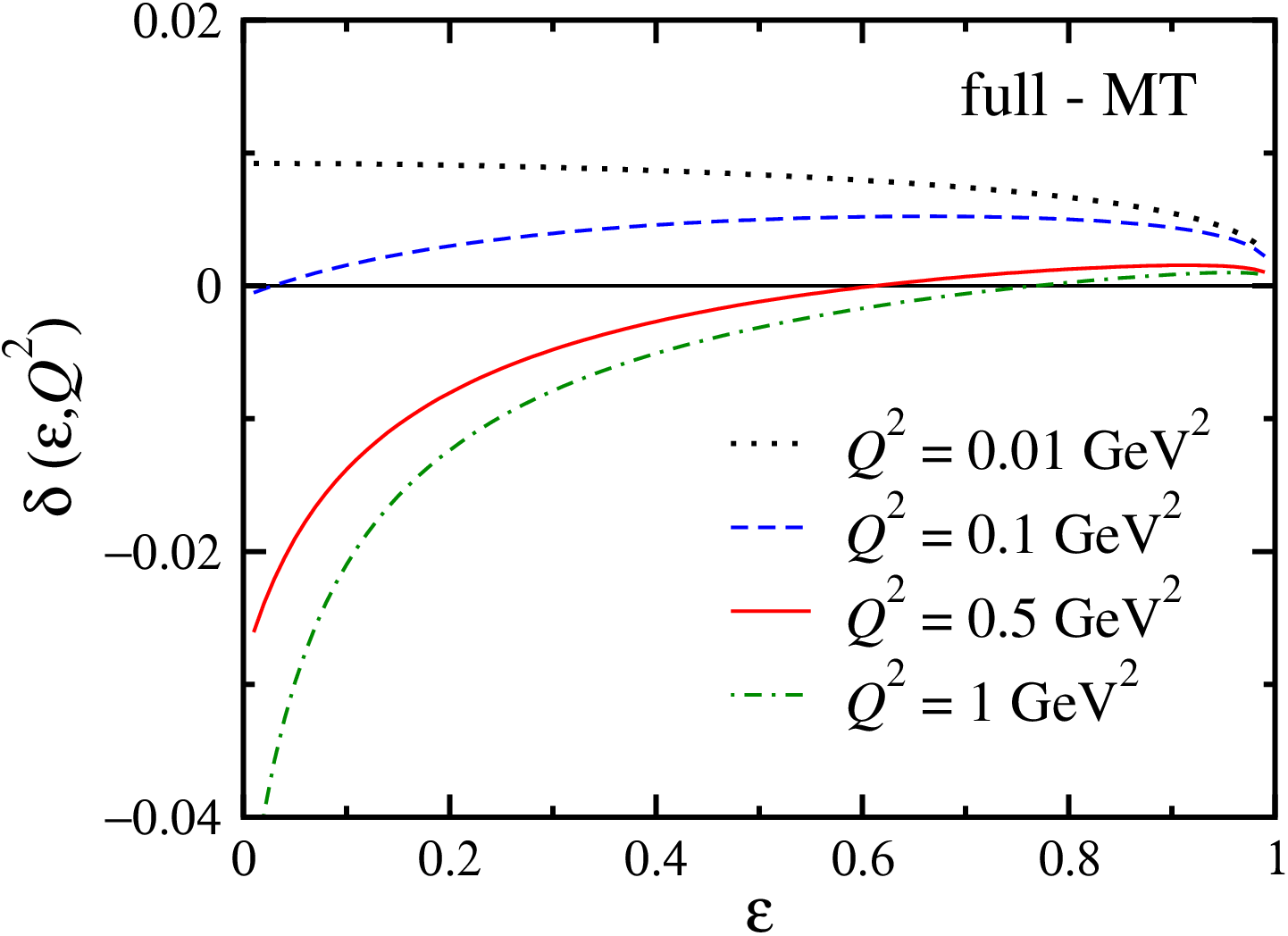}
\includegraphics[height=6cm]{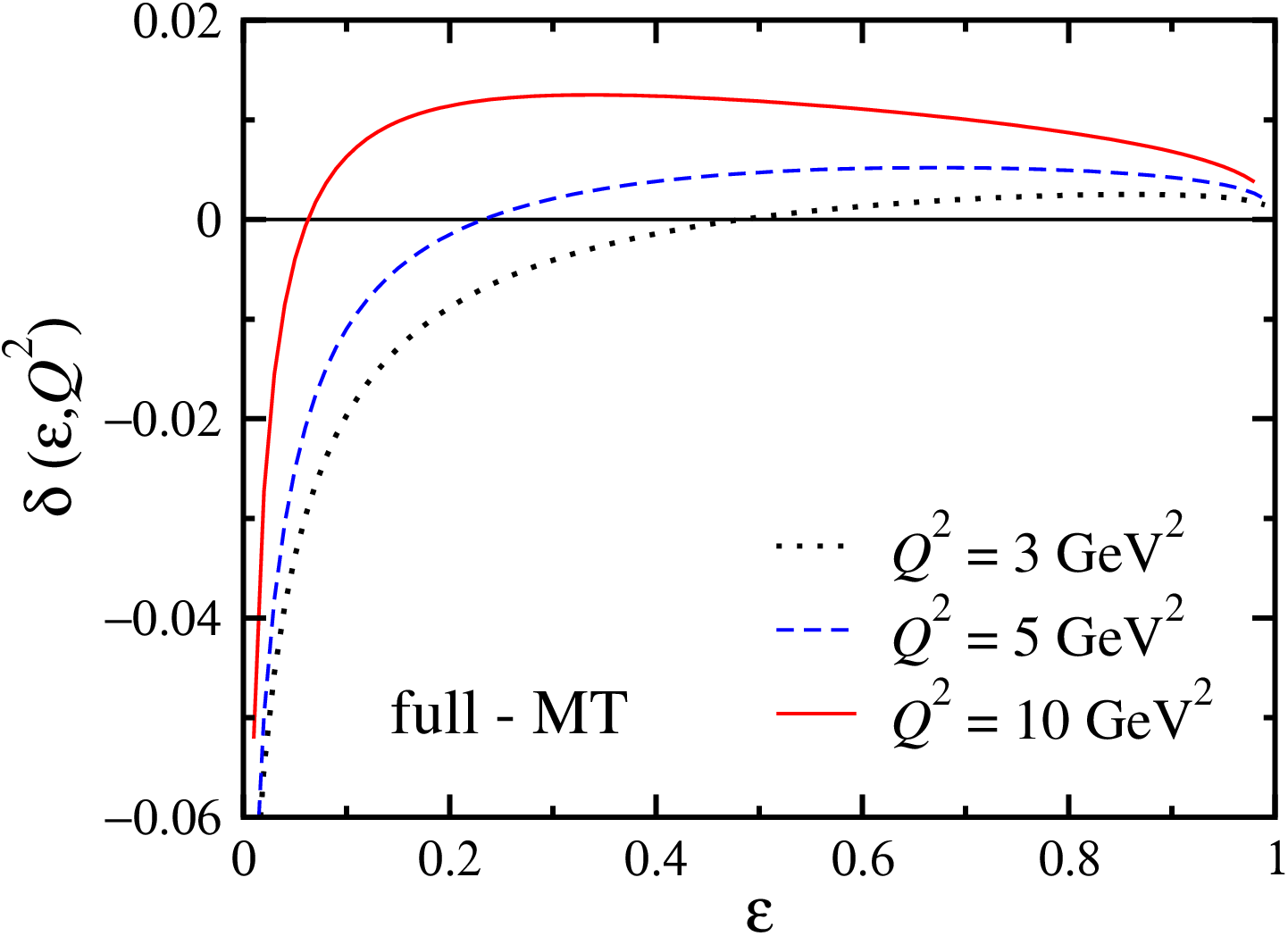}
\caption{(Color online)
	Two-photon exchange correction to the pion form factor squared
	as a function of $\varepsilon$ for various $Q^2$,
	relative to the Mo \& Tsai (MT) \cite{MT69} contribution.
	A monopole parametrization is used for $F_\pi$ in the
	full calculation. \\}
\label{fig:epsMT}
\end{figure}

To quantify the effect of the IR-finite, hadron structure dependent
contribution, in Fig.~\ref{fig:epsMT} we show the difference between
the full TPE correction $\delta$ and the MT prescription \cite{MT69}
as a function of $\varepsilon$ for various $Q^2$.
In the numerical calculations we use a monopole parametrization
for the ``bare'' pion form factor in Eq.~(\ref{eq:Mgg}),
\begin{eqnarray}
F_\pi(q^2) &=& {1 \over \left( 1 - q^2/\Lambda^2 \right)}\, ,
\end{eqnarray}
with $\Lambda = 770$~MeV corresponding to the $\rho$-meson mass.
The loop integrals of Eq.~(\ref{eq:Mgg}) can then be done analytically,
and expressed in terms of Passarino-Veltman 2-, 3-, and 4-point
functions \cite{HV79}.  In the calculations we use the computer
program FEYNCALC \cite{FC91}.

At low $Q^2$ ($Q^2 \sim 0.01$~GeV$^2$) the TPE correction is positive
and of the order of 1\% at backward angles (small $\varepsilon$),
decreasing to zero in the $\varepsilon \to 1$ (forward angle) limit.
With increasing $Q^2$ the correction becomes smaller (more negative) up
to $Q^2 \sim 1-2$~GeV$^2$, especially in the extreme backward region
($\varepsilon \to 0$), but changes sign at intermediate $\varepsilon$.
Note, however, that unlike electron-proton scattering, the electron-pion
scattering cross section vanishes at the extreme backward angles limit
($\varepsilon = 0$).  Above $Q^2 \sim 2$~GeV$^2$ the correction grows
once again, reaching $\sim 1\%$ at $Q^2 = 10$~GeV$^2$.

\begin{figure}[t]
\includegraphics[height=6cm]{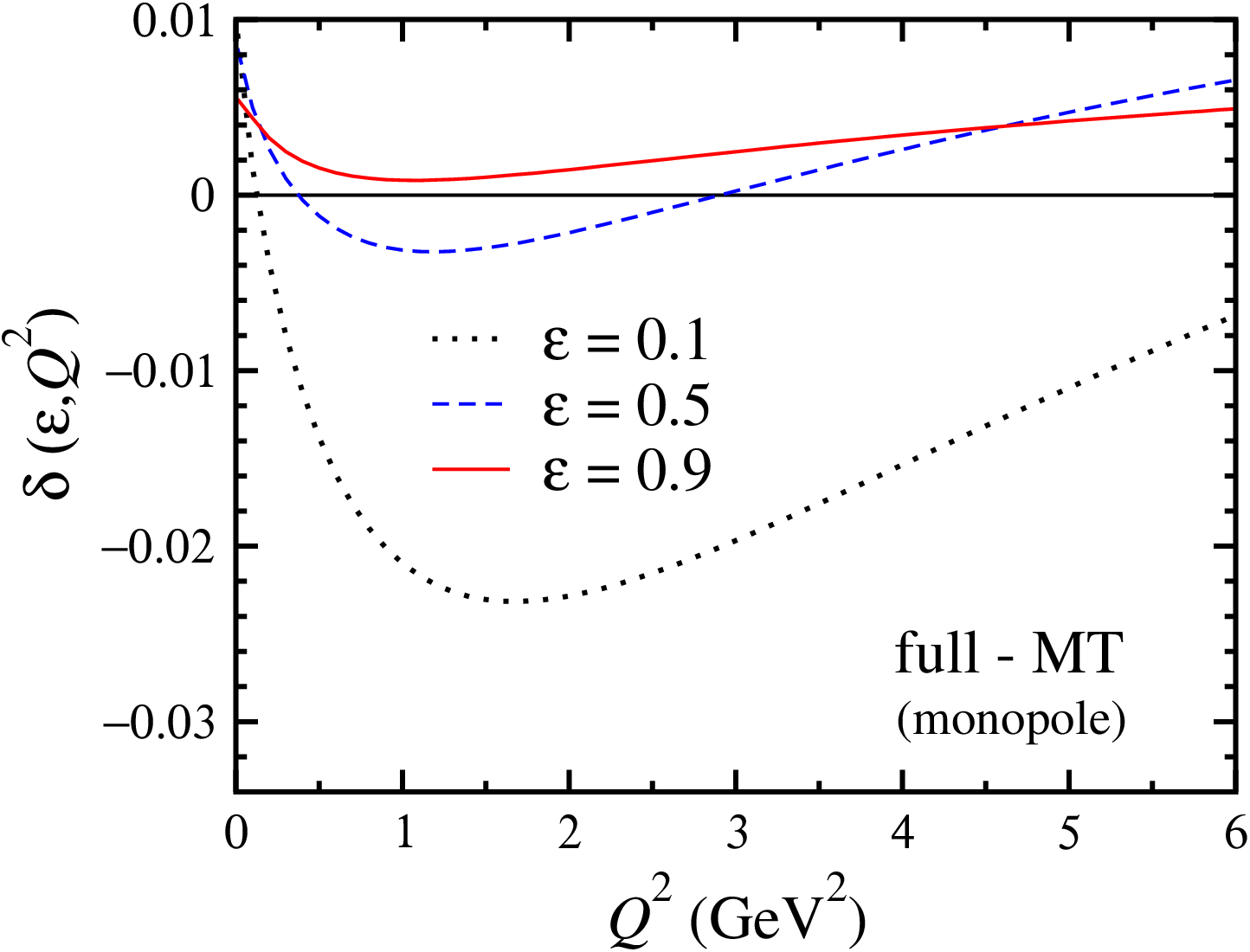}
\includegraphics[height=6cm]{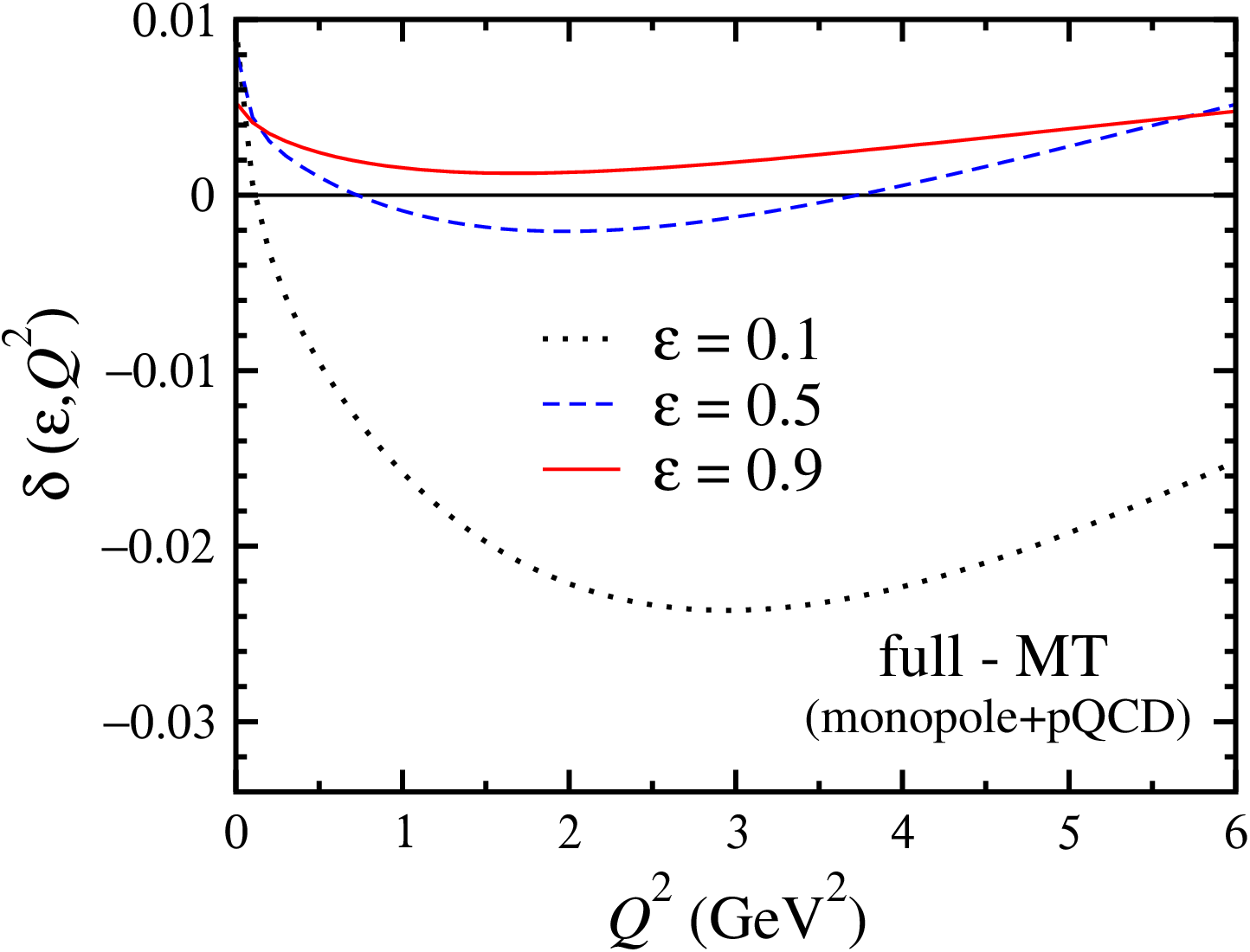}
\caption{(Color online)
        Two-photon exchange correction to the pion form factor
	squared as a function of $Q^2$ for various $\varepsilon$,
	relative to the Mo \& Tsai (MT) \cite{MT69} contribution,
	for the monopole (left) and monopole $+$ pQCD (right)
	parametrizations of $F_\pi(q^2)$. \\}
\label{fig:qMT}
\end{figure}

The $Q^2$ dependence is more clearly illustrated in Fig.~\ref{fig:qMT},
where $\delta$ is shown for fixed $\varepsilon$ over the range
$Q^2 = 1-6$~GeV$^2$.  Interestingly, the correction is most positive
at very small $Q^2 \ll 1$~GeV$^2$ and large $Q^2 \gg 1$~GeV$^2$,
reaching its minimum values at $Q^2 \sim 1-2$~GeV$^2$.  At small
$\varepsilon$ the $Q^2$ dependence is seen to change most rapidly.

While the monopole parametrization is known to give a good description
of the pion form factor data at low $Q^2$, it overestimates $F_\pi(q^2)$
at larger $Q^2$.  An alternative parametrization to the monopole which
fits the available data and builds in gauge invariance constraints for
the $Q^2 \to 0$ limit and perturbative QCD expectations for the
$Q^2 \to \infty$ behavior was given in Ref.~\cite{pidual}.
Using this parametrization the TPE correction $\delta$ is shown in
Fig.~\ref{fig:qMT} (right panel).  As expected, the differences at
low $Q^2$ are negligible, but become noticeable at high $Q^2$.
The qualitative behavior of the corrections, however, is not affected
by the specific form chosen.

\begin{figure}[t]
\includegraphics[height=6cm]{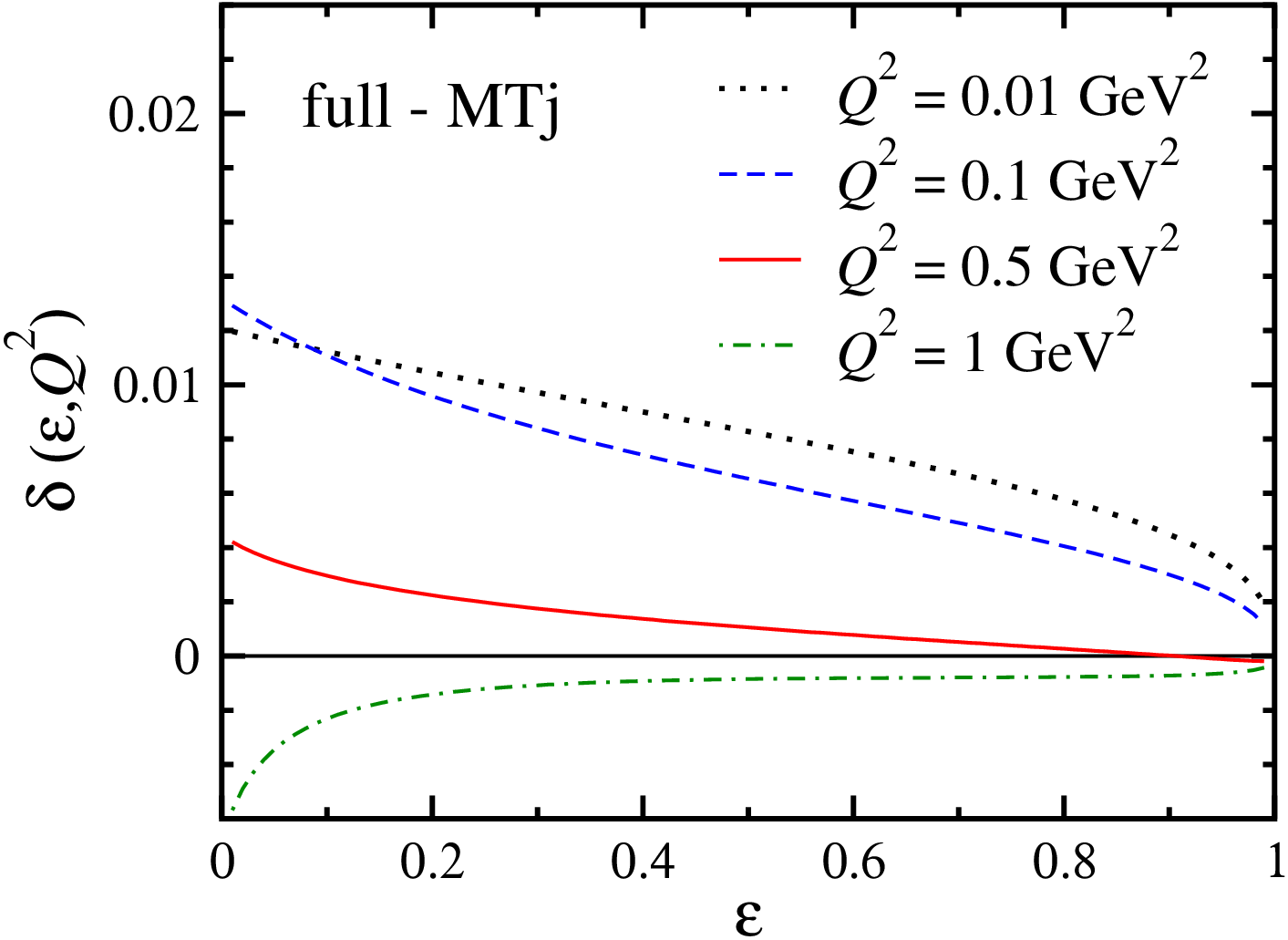}
\includegraphics[height=6cm]{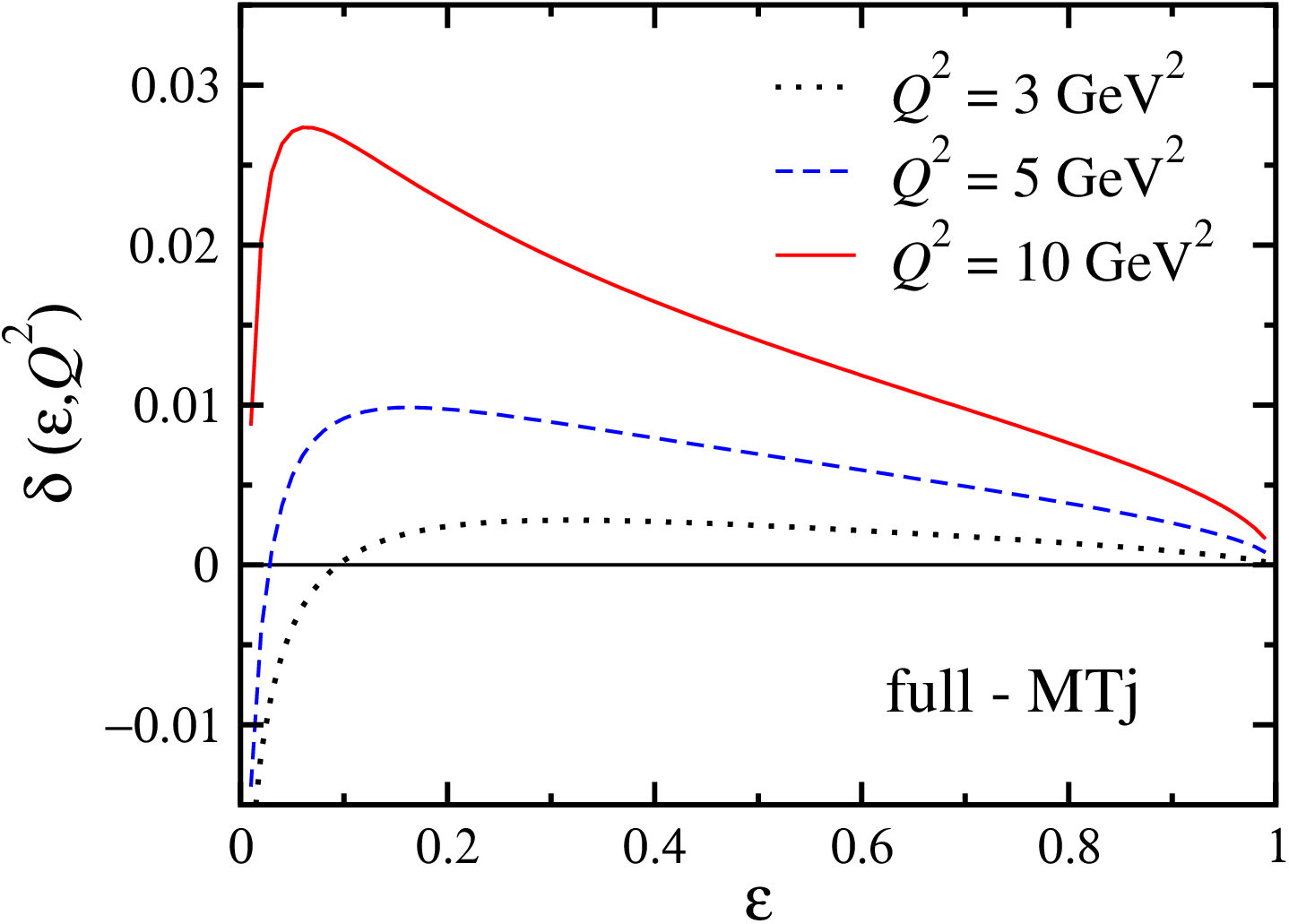}
\caption{(Color online)
        Two-photon exchange correction to the pion form factor squared
	as a function of $\varepsilon$ for various $Q^2$, relative
	to the Maximon \& Tjon (MTj) \cite{MTj} contribution.
	A monopole parametrization is used for $F_\pi$ in the
        full calculation. \\}
\label{fig:epsMTj}
\end{figure}

We should note that the effects illustrated in Figs.~\ref{fig:epsMT}
and \ref{fig:qMT} are not physical, but merely reflect the accuracy
with which the full result can be approximated by the particular
prescription for the IR-divergent contribution.  It is only relevant
because the MT approximation is widely used for applications of
radiative corrections in analyses of electron scattering data
\cite{FpiJLab,Jones,Ent}.
An alternative prescription was suggested by Maximon \& Tjon (MTj)
\cite{MTj}, who also approximate the TPE by the contributions at the
photon poles, setting $k_1\to 0$ and $k_2\to 0$ in the numerator of
Eq.~(\ref{eq:Mgg}).  However, no further approximation is made in
evaluating the remaining loop integration, which can be written
in terms of 4-point functions.  In the limit $Q^2 \gg m_e^2$ the
4-point functions simplify significantly, and the TPE correction
$\delta_{\rm IR}^{(\rm MTj)}$ can be written as
\begin{eqnarray}
\delta_{\rm IR}^{(\rm MTj)}
&=& = - {2 \alpha \over \pi} \ln{ \left( E_1 \over E_3 \right) }
        \ln{ \left( Q^2 \over \lambda^2 \right) }\, .
\label{eq:IR_MTj}
\end{eqnarray}
As before, the logarithmic $\lambda$ singularity is cancelled
exactly by the inelastic bremsstrahlung contribution.

\begin{figure}[t]
\includegraphics[height=6cm]{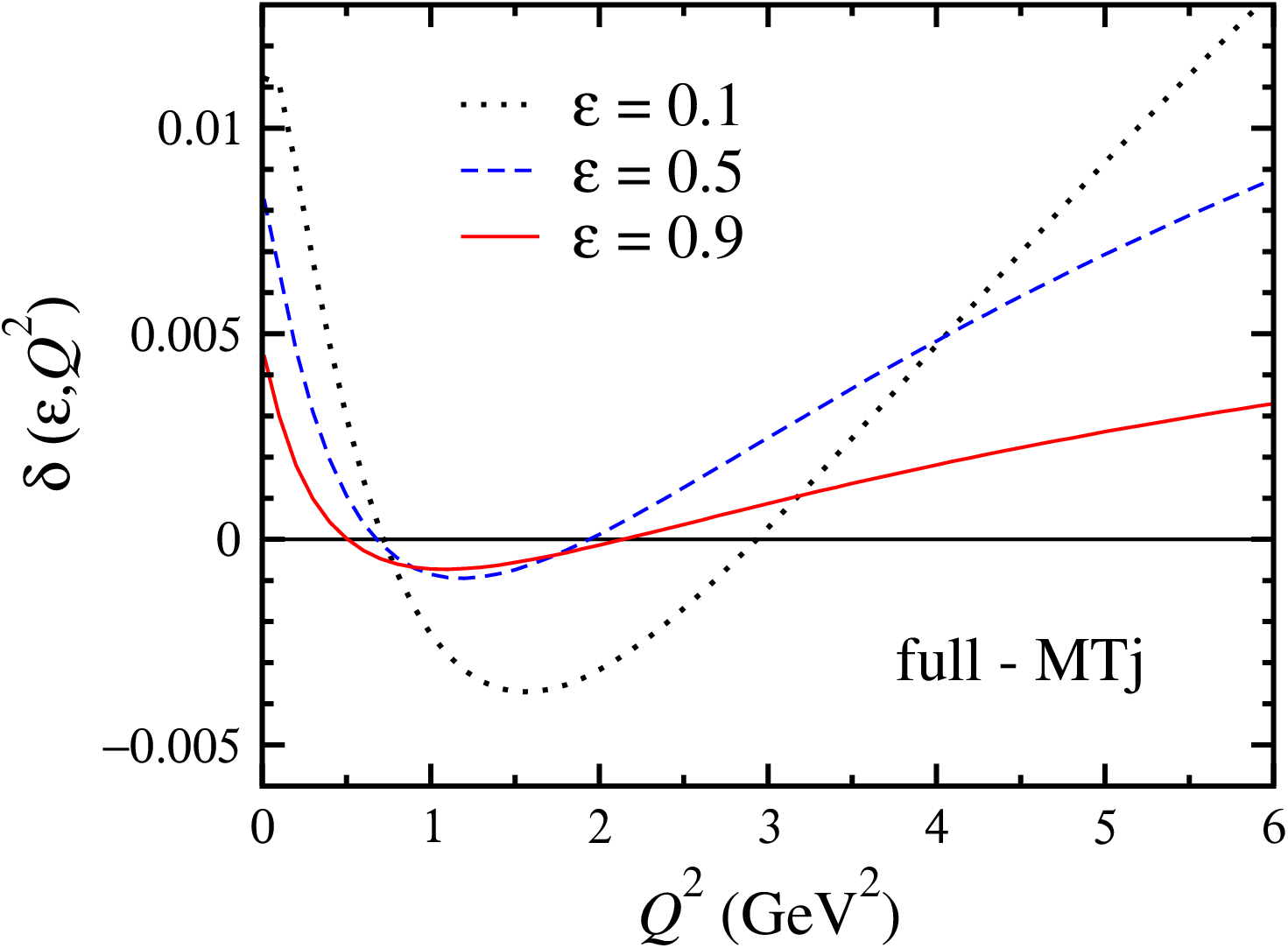}
\caption{(Color online)
        Two-photon exchange correction to the pion form factor squared
	as a function of $Q^2$ for various $\varepsilon$, relative
	to the Maximon \& Tjon (MTj) \cite{MTj} contribution.
	A monopole parametrization is used for $F_\pi$ in the
        full calculation. \\}
\label{fig:qMTj}
\end{figure}

The difference between the full, hadron structure dependent
result for $\delta$ and the MTj approximation is illustrated
in Fig.~\ref{fig:epsMTj} as a function of $\varepsilon$.
At low $Q^2$ ($Q^2 \lesssim 0.1$~GeV$^2$) the correction is
similar to that in Fig.~\ref{fig:epsMT} for the MT IR prescription,
but is generally smaller in magnitude at larger $Q^2$.
In particular, it displays a somewhat milder $\varepsilon$
dependence for backward angles at higher $Q^2$.
Qualitatively, however, the behavior of the correction as a
function of $Q^2$ is similar for the MTj and MT prescriptions,
as Fig.~\ref{fig:qMTj} shows, with $\delta$ most positive at
low $Q^2$, $Q^2 \ll 1$~GeV$^2$ (for all $\varepsilon$),
decreasing to a minimum at $Q^2 = 1-2$~GeV$^2$, before rising
again for larger $Q^2$.

Note that in contrast to the proton form factor case, where the TPE
effects give large corrections to the elastic form factors extracted
from longitudinal-transverse (LT) separated cross sections at large
$Q^2$ \cite{AMT}, the TPE corrections to the pion form factor are
relatively small.
This is related to the fact that electron scattering from a spin-0
target is described by a single form factor, with no LT separation
necessary.
On the other hand, since $F_\pi$ is extracted by performing an LT
separation of the pion electroprodution cross section, TPE with one
photon attached to the pion and the other to the initial proton or
final neutron could modify the longitudinal cross section, and may
need to be considered.

The reliability of the results at high $Q^2$ ($Q^2 \gg 1$~GeV$^2$)
may be more questionable since only pion elastic intermediate states
are included in this analysis, and we expect excited hadronic states
to play a greater role with increasing $Q^2$ \cite{KBMT05,KB07}.
However, since the mass difference between the pion and the next
excited resonant state, the $\rho$ meson, is almost 5 times as large
as the pion mass, we would not expect these contributions to be
significant for the pion.  The nonresonant contributions and the
off-shell dependence of the pion form factor, on the other hand,
may need to be examined in future analyses.

In summary, we have computed the two-photon exchange contribution
to the pion electromagnetic form factor arising from the finite
size of the pion.  Compared with the standard infrared contribution
computed in the soft-photon approximation, the hadron structure
dependent corrections are $\lesssim 1\%$ at low $Q^2$
($Q^2 \sim 0.01$~GeV$^2$), but increase to several percent at
$Q^2 \gtrsim 1$~GeV$^2$ at backward angles.  These contributions
will need to be taken into account in the treatment of radiative
corrections for extractions of the pion form factor from
high-precision pion electroproduction data.

\acknowledgments

We thank J.~Arrington, D.~Gaskell and G.~Huber for helpful
discussions.
P.~G.~B. thanks the Theory Center at Jefferson Lab for support
during a sabbatical leave, where this work was performed.
This work was supported in part by NSERC (Canada), DOE grant
DE-FG02-93ER-40762, and DOE contract DE-AC05-06OR23177, under
which Jefferson Science Associates, LLC operates Jefferson Lab.



\begin{thebibliography}{99}

\bibitem{FpiJLab}
T.~Horn {\it et al.},
Phys.\ Rev.\ Lett.\  {\bf 97}, 192001 (2006);
%
V.~Tadevosyan {\it et al.},
Phys.\ Rev.\  C {\bf 75}, 055205 (2007);
%
G.~M.~Huber {\it et al.},
Phys.\ Rev.\  C {\bf 78}, 045203 (2008).

\bibitem{Jones}
M.~K.~Jones {\em et al.},
Phys.\ Rev.\ Lett.\ {\bf 84}, 1398 (2000);
O.\ Gayou {\em et al.},
Phys.\ Rev.\ Lett.\ {\bf 88}, 092301 (2002);
%
V.~Punjabi {\em et al.},
Phys.\ Rev.\ C {\bf 71}, 055202 (2005)
[Erratum-ibid.\ C {\bf 71}, 069902 (2005)].

\bibitem{review}
C.~E.~Carlson and M.~Vanderhaeghen,
Ann.\ Rev.\ Nucl.\ Part.\ Sci.\  {\bf 57}, 171 (2007)
%
J.~Arrington, P.~G.~Blunden, W.~Melnitchouk and J.~A.~Tjon,
in preparation for Prog. Nucl. Part. Phys. (2009).

\bibitem{BMT03}
P.~G.~Blunden, W.~Melnitchouk and J.~A.~Tjon,
Phys.\ Rev.\ Lett.\ {\bf 91}, 142304 (2003).

\bibitem{BMT05}
P.~G.~Blunden, W.~Melnitchouk and J.~A.~Tjon,
Phys.\ Rev.\  C {\bf 72}, 034612 (2005).

\bibitem{KBMT05}
S.~Kondratyuk, P.~G.~Blunden, W.~Melnitchouk and J.~A.~Tjon,
Phys.\ Rev.\ Lett.\  {\bf 95}, 172503 (2005).

\bibitem{KB07}
S.~Kondratyuk and P.~G.~Blunden,
Phys.\ Rev.\  C {\bf 75}, 038201 (2007).

\bibitem{MT69} 
L.~W.~Mo and Y.~S.~Tsai, 
Rev.\ Mod.\ Phys.\ {\bf 41}, 205 (1969);
Y.~S.~Tsai,
Phys.\ Rev.\ {\bf 122}, 1898 (1961).

\bibitem{HV79}
G.~'t~Hooft and M.~Veltman,
Nucl.\ Phys.\ {\bf B153}, 365 (1979).

\bibitem{FC91}
R.~Mertig, M.~B\"ohm and A.~Denner,
Comput.\ Phys.\ Commun.\ {\bf 64}, 345 (1991).

\bibitem{pidual}
W.~Melnitchouk,
Eur. Phys. J. A {\bf 17}, 223 (2003).

\bibitem{Ent}
R.~Ent {\it et al.},
Phys.\ Rev.\  C {\bf 64}, 054610 (2001).

\bibitem{MTj}
L.~C.~Maximon and J.~A.~Tjon,
Phys.\ Rev.\ C {\bf 62}, 054320 (2000).

\bibitem{AMT}
J.~Arrington, W.~Melnitchouk and J.~A.~Tjon,
Phys.\ Rev.\  C {\bf 76}, 035205 (2007).

\end{thebibliography}
\end{document}